\documentclass[english]{elsarticle}
\usepackage[T1]{fontenc}
\usepackage[latin9]{inputenc}
\usepackage{geometry}
\usepackage{float}
\usepackage{amsmath}
\usepackage{amssymb}

\makeatletter

\providecommand{\tabularnewline}{\\}

\usepackage{easybmat}

\makeatother

\usepackage{babel}
\begin{document}

\section*{Model}

\subsection*{Evolution of Voltage over Time}

Consider a group of interacting populations of neurons. Equation \ref{eq:single_pop_def}
describes the evolution over time of the membrane voltages for the
$i\textrm{th}$ ensemble. Table \ref{var-explan} explains the remaining
variables.

\begin{equation}
\tau_{i}\frac{d\vec{v_{i}}}{dt}=-\vec{v}_{i}+\mathbf{M}_{ii}\vec{v}_{i}+\sum_{i\neq j}\mathbf{M}_{ij}\vec{v}_{j}+\mathbf{W}_{i}\vec{u}\label{eq:single_pop_def}
\end{equation}

\begin{table}[H]
\centering%
\begin{tabular}{cl}
Symbol & Meaning\tabularnewline
$\mathbf{M}_{ij}$ & Connections from $j\textrm{th}$ ensemble to $i\textrm{th}$ \tabularnewline
$\mathbf{W}_{i}$ & Weight of input for $i\textrm{th}$ ensemble\tabularnewline
$\vec{u}$ & Input\tabularnewline
$\tau$ & Time constant\tabularnewline
\end{tabular}

\caption{Meaning of symbols in Equation \ref{eq:single_pop_def}}
\label{var-explan}

\end{table}

To represent $n$ ensembles, we may combine $n$ versions of Equation
\ref{eq:single_pop_def}, as Equation \ref{eq:no-subs} describes.

\begin{equation}
\tau_{v}\frac{d\vec{v}}{dt}=-\vec{v}+\mathbf{M}\vec{v}+\mathbf{W}\vec{u}\label{eq:no-subs}
\end{equation}

\[
\vec{v}=\left(\begin{array}{c}
\vec{v_{i}}\\
\vec{v}_{j}\\
\vdots\\
\vec{v_{n}}
\end{array}\right),\;\mathbf{M}=\begin{bmatrix}\mathbf{M}_{ii} & \mathbf{M}_{ij} & \ldots & \mathbf{M}_{in}\\
\mathbf{M}_{ji} & \mathbf{M}_{jj}\\
\vdots &  & \ddots\\
\mathbf{M}_{ni} &  &  & \mathbf{M}_{nn}
\end{bmatrix}
\]

If the real part of the eigenvalues of $\mathbf{M}$ are less than
one, then the system will evolve to the voltage that Equation \ref{eq:v_steady}
describes. 

\begin{equation}
\vec{v}_{\infty}=\mathbf{KW}\vec{u}\quad\mathbf{K}=\left(\mathbf{I}-\mathbf{M}\right)^{-1}\label{eq:v_steady}
\end{equation}

\subsection*{Islands of Steady State Behavior}

$\mathbf{M}$ need not be symmetric. But, if some block matrices within
it are, the corresponding subpopulations can approach their own steady
state even if the network is still unstable. If, furthermore, the
combination of recurrent and feedforward input to those subpopulations
is a saturating function, then the system's Lyapunov function is bounded
and fixed points for that subpopulation \emph{must }exist (Cohen and
Grossberg, 1983). 

One reasonable way to create a symmetric weight matrix for $\mathbf{M}_{k}$,
the block matrix that describes the $k\textrm{th}$ ensemble is to
assume that it recognizes one of $N$ memory patterns, $\left\{ v^{1},v^{2},\ldots,v^{m},\ldots,v^{N}\right\} $.
Assume that a subpopulation with $n$ neurons signals its recognition
of any memory pattern, $\vec{v}^{m}$, by displaying a voltage vector
$c\vec{v}^{m}$. One matrix that accomplishes this is: 

\begin{equation}
\mathbf{M}=\frac{\lambda}{c^{2}\alpha N\left(1-\alpha\right)}\sum^{N}\left(\vec{v}-\alpha cn\right)\otimes\left(\vec{v}-\alpha cn\right)-\frac{n\otimes n}{\alpha N}\label{eq:assoc-mem}
\end{equation}

\subsection*{Longer-term Effects on Neural Population Dynamics}

Assume that two additional processes occur as defined by Equations
\ref{eq:oja} and \ref{eq:godall}. Both are much slower than Equation
\ref{eq:no-subs}. One modifies $\mathbf{W}\textrm{ and an even slower one modifies }\mathbf{M}$.

\begin{equation}
\tau_{W}\frac{d\mathbf{W}}{dt}=\left\langle \vec{v}\vec{u}\right\rangle -\alpha\left\langle \vec{v}\vec{v}\right\rangle \mathbf{W}\quad\alpha>0\label{eq:oja}
\end{equation}

\begin{equation}
\tau_{M}\frac{d\mathbf{M}}{dt}=-\left(\mathbf{W}\vec{u}\right)\vec{v}+\mathbf{K}^{-1}\label{eq:godall}
\end{equation}

Replacing $\vec{v}\textrm{ with }\vec{v}_{\infty}$in Equations \ref{eq:oja}
and \ref{eq:godall} yields Equations \ref{eq:oja_2} and \ref{eq:godall_2}.
Note the appearance of the autocorrelation matrices for the stimulus,
$\mathbf{Q}$, and network activity, $\mathbf{R}$.

\begin{equation}
\tau_{W}\frac{d\mathbf{W}}{dt}=\mathbf{KWQ}-\alpha\mathbf{R_{\infty}}\mathbf{W}\quad\mathbf{Q}=\left\langle \vec{u}\vec{u}\right\rangle ,\,\mathbf{R_{\infty}}=\left\langle \vec{v_{\infty}}\vec{v_{\infty}}\right\rangle \label{eq:oja_2}
\end{equation}

\begin{equation}
\tau_{M}\frac{d\mathbf{M}}{dt}=-\left(\mathbf{W}\vec{u}\right)\left(\mathbf{KW}\vec{u}\right)+\mathbf{K}^{-1}\label{eq:godall_2}
\end{equation}

Equations \ref{eq:oja_2} and \ref{eq:godall_2} describe somewhat
contrasting behaviors. Equation \ref{eq:oja_2} aligns the correlation
structure of the network activity with the strength of its recurrent
connections. Equation \ref{eq:godall_2} urges the outputs to be decorrelated.
A faster correlating influence and slower decorrelating one allow
oscillations in the correlation of network activity. According to
this model, interestingly, those oscillations are dependent on input
but not directly on the correlation structure of the input.

\section*{Remarks on the Structure of the Model}

There is an interesting concordance between $\mathbf{R}$ and Equation
\ref{eq:assoc-mem}. If $\mathbf{R}_{mem}=\left\langle \vec{v}_{mem}\vec{v}_{mem}\right\rangle $,
then $\mathbf{M}\textrm{ }\varpropto\sum\mathbf{R}_{mem}$. This further
highlights how interrelated Equations \ref{eq:oja_2} and \ref{eq:godall_2}
are.

\subsection*{Effect of Correlation Structure on the Dynamics of the Feedforward
Weights}

If the stimulus is random, that is $\mathbf{Q}=\mathbf{I}$, then
the feedforward weights, $\mathbf{W}$, stop changing only when $\mathbf{K}=\alpha\mathbf{R}_{\infty}$.
Combining the definition of $\mathbf{K}$ in Equation \ref{eq:v_steady}
with the observation that $\mathbf{R}_{\infty}$ must have$\textrm{rank }1$
we note that such stimuli prevent this system from recognizing any
pattern that requires a distribution of activities over the network.
Moreover, considering the rank-nullity theorem and ranks of $\mathbf{Q}\textrm{ and }\mathbf{R}$,
one can see that this result holds for any stimulus autocorrelation
matrix, $\mathbf{Q}$, that results from the outer product of a vector
with itself. 

It is next natural to consider how this system responds to many superimposed
stimuli that each have different correlation structures. That is,
consider a$\mathbf{Q}$ that results from the sum of $i$ correlation
matrices. Each of those autocorrelation matrices results from the
outer product of the $i\textrm{th}$ activity pattern with itself,
as Equation \ref{eq:multi-q} describes. 

\begin{equation}
\mathbf{Q}=\sum_{i}\left\langle \vec{u}_{i}\vec{u}_{i}\right\rangle \label{eq:multi-q}
\end{equation}

If we assume that all the input patterns are pairwise independent,
then the rank of $\mathbf{Q}$ becomes equal to $i$, whose upper
bound we assume to be the number of neurons. This stands in contrast
to the result of the previous paragraph.

\subsection*{Effects of Correlation Structure on the Dynamics of Network Connections}

By studying Equation \ref{eq:godall_2} we can gain some insight into
how $\vec{u}\textrm{ and }\mathbf{W}$ influence $\mathbf{M}$. If
$\vec{u}$ lies in the nullspace of the columns of $\mathbf{W}$,
then $\mathbf{M}$ approaches the identity matrix. Said another way,
when $\vec{u}\textrm{ and }\mathbf{W}$ jam each other,$\mathbf{M}$
tries to span the largest basis it can. 

The contrary effects of

\subsection*{Nonrandom Constant Stimulus}

Consider any brief nonrandom stimulus such that, $\mathbf{Q}\neq\mathbf{I}$
but is constant. Then$\mathbf{W}$ only stops changing if the columns
of $\mathbf{Q}$ are in the nullspace of the rows of $\mathbf{W}$.
Because$\mathbf{Q}$ results from the tensor product of a vector with
itself, it has $\textrm{rank }1$ and so it may lie in the nullspace
of $\mathbf{W}$.

\subsection*{Brief Pulse }

If a stimulus is presented to the system for a brief period of time,
it will only cause sustained activity in the system if the activity
that it induces is itself a fixed point of the system. That is, for
a stimulus, $\vec{u}$, and its response, $\vec{v}$, Equation must
hold.

\subsection*{Extension to Drug Addiction}

\paragraph*{Craving }

\paragraph*{Similar response to variable reinforcement and huge rewards}

Let the $i\textrm{th}$ subpopulation of Equation \ref{eq:no-subs},
in analogy with a proposed role for the dopaminergic neurons in ventral
tegmental area (VTA) represent the positive expecrted reward of a
stimulus. Let another population, the $j\textrm{th}$ one, in analogy
with the GABAergic neurons in the VTA represent the negative expected
reward. 
\end{document}